\documentclass[usenatbib,usegraphicx]{mn2e}
\usepackage{amsfonts}
\usepackage{amsmath}
\usepackage{amssymb}
\usepackage{comment}
\usepackage{float}
\usepackage[usenames,dvipsnames]{color}

\title[The Horizon-AGN simulation]
{The Horizon-AGN simulation: evolution of galaxy properties over
cosmic time}
\author[S. Kaviraj et al.]
{S. Kaviraj,$^{1}$\thanks{s.kaviraj@herts.ac.uk} C.
Laigle,$^{2}$ T. Kimm,$^{3}$ J. E. G. Devriendt,$^{4,5,6}$ Y. Dubois,$^{2}$  \newauthor C. Pichon,$^{2,7}$A. Slyz,$^{4}$ E. Chisari,$^{4}$ and S. Peirani$^{2,8}$\\
$^{1}$Centre for Astrophysics Research, University of
Hertfordshire, College Lane, Hatfield, Herts, AL10 9AB, UK\\
$^{2}$Institut d'Astrophysique de Paris, Sorbonne Universit\'es,
UPMC Univ Paris 06 et CNRS, UMR 7095, 98 bis bd Arago, 75014
Paris, France\\
$^{3}$Institute of Astronomy, Madingley Road, Cambridge, CB3 0HA, UK\\
$^{4}$Dept of Physics, University of Oxford, Keble Road, Oxford
OX1 3RH UK\\
$^{5}$CRAL, CNRS, Universit\'e Claude Bernard Lyon I, ENS Lyon,
UMR 5574\\
$^{6}$Observatoire de Lyon, 9 Avenue Charles Andr\'e, F-69230
Saint-Genis-Laval, France\\
$^{7}$Korea Institute of Advanced Studies (KIAS) 85 Hoegiro,
Dongdaemun-gu, Seoul, 02455, Republic of Korea\\
$^{8}$Kavli Institute for the Physics and Mathematics of the
Universe, The University of Tokyo, Kashiwa, Chiba 277-8583, Japan}

\begin{document}

\maketitle

\def \aj {AJ}
\def \mnras {MNRAS}
\def \pasp {PASP}
\def \apj {ApJ}
\def \apjs {ApJS}
\def \apjl {ApJL}
\def \aap {A\&A}
\def \nat {Nature}
\def \araa {ARAA}
\def \iaucirc {IAUC}
\def \aaps {A\&A Suppl.}
\def \qjras {QJRAS}
\def \na {New Astronomy}
\def \aapr {A\&ARv}
\def\lesssim{\mathrel{\hbox{\rlap{\hbox{\lower4pt\hbox{$\sim$}}}\hbox{$<$}}}}
\def\gtrsim{\mathrel{\hbox{\rlap{\hbox{\lower4pt\hbox{$\sim$}}}\hbox{$>$}}}}


\begin{abstract}
We compare the predictions of Horizon-AGN, a hydro-dynamical
cosmological simulation that uses an adaptive mesh refinement
code, to observational data in the redshift range $0<z<6$. We
study the reproduction, by the simulation, of quantities that
trace the aggregate stellar-mass growth of galaxies over cosmic
time: luminosity and stellar-mass functions, the star formation
main sequence, rest-frame UV-optical-near infrared colours and the
cosmic star-formation history. We show that Horizon-AGN, which is
not tuned to reproduce the local Universe, produces good overall
agreement with these quantities, from the present day to the epoch
when the Universe was 5\% of its current age. By comparison to
Horizon-noAGN, a twin simulation without AGN feedback, we quantify
how feedback from black holes is likely to help shape galaxy
stellar-mass growth in the redshift range $0<z<6$, particularly in
the most massive galaxies. Our results demonstrate that
Horizon-AGN successfully captures the evolutionary trends of
observed galaxies over the lifetime of the Universe, making it an
excellent tool for studying the processes that drive galaxy
evolution and making predictions for the next generation of galaxy
surveys.
\end{abstract}




\section{Introduction}
In recent years, an explosion of multi-wavelength survey data has
enabled us to probe the evolution of galaxy properties over
$\sim$90\% of cosmic time. To interpret these observations in the
context of the $\Lambda$CDM paradigm
\citep[e.g.][]{Rees1977,White1978a}, and understand the processes
that underpin galaxy formation, well-calibrated models are
required, that reproduce the broad evolution of galaxy properties
over the lifetime of the Universe.

In this paradigm, initial density perturbations, that are
gravitationally amplified, collapse to form dark matter halos.
Smaller halos form first and merge under the influence of gravity
to form progressively larger ones
\citep[e.g.][]{Peebles1982,Blumenthal1984}. Cold gas settles into
the dark-matter potential wells within rotationally-supported
discs \citep[e.g.][]{Fall1980,Mo1998}, with star formation
regulated by the local gas density
\citep[e.g.][]{Schmidt1959,Kennicutt1998}. Supernovae enrich the
inter-stellar medium with metals, and inject thermal and kinetic
energy into the ambient gas \citep[e.g.][]{Scannapieco2008}.
Further sources of energetic feedback are active galactic nuclei
(AGN), which release part of the rest-mass energy of accreted
matter into the gas reservoir of their host galaxies
\citep[e.g.][]{Fabian2012}.

\begin{figure*}
\begin{minipage}{172mm}
\begin{center}
$\begin{array}{c}
\includegraphics[width=\textwidth]{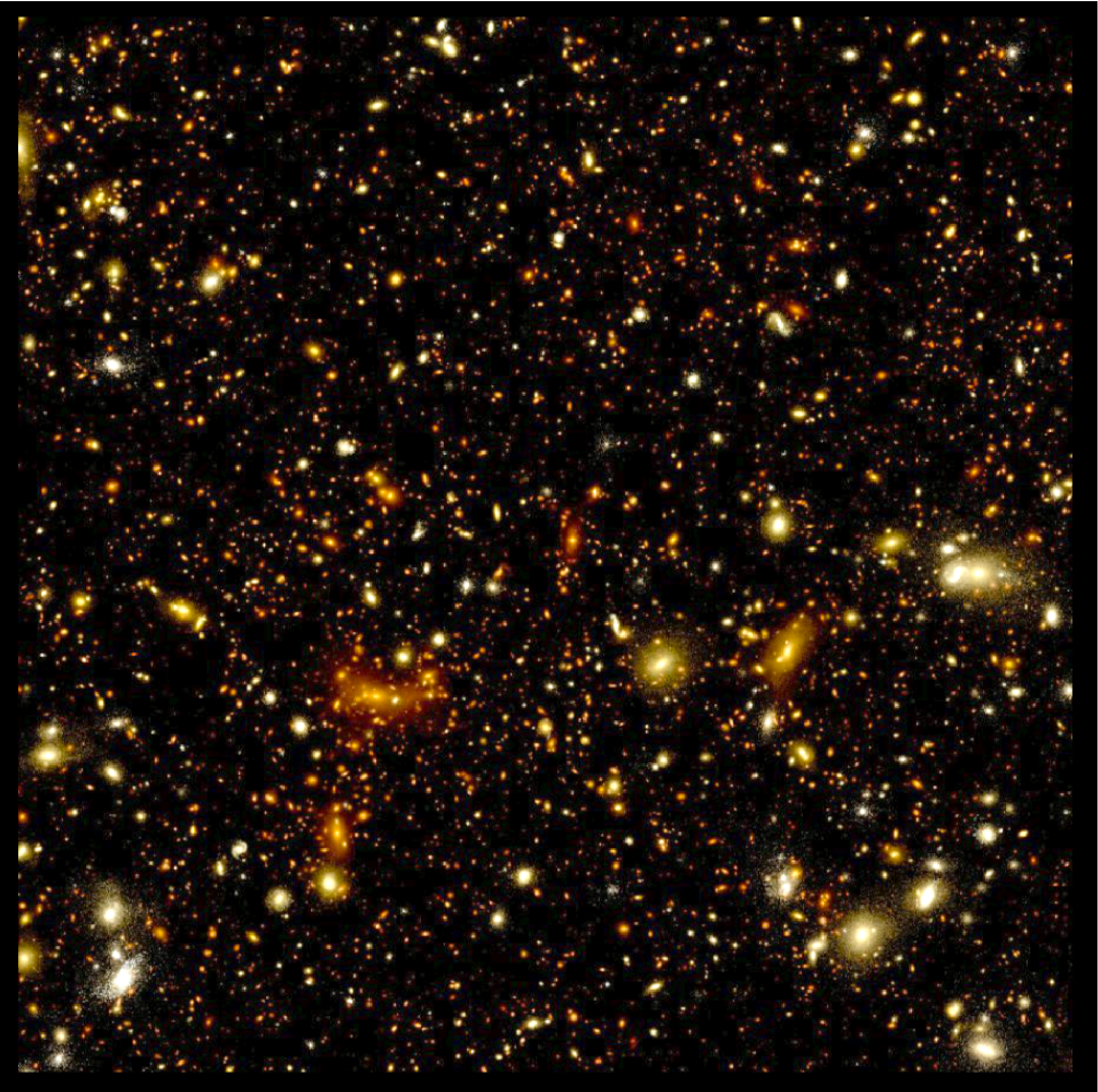}
\end{array}$
\caption{{\color{black}A 14 arcmin$^2$ simulated composite image
from the Horizon-AGN lightcone \citep{Pichon2010}, in the $u$, $r$
and $z$ filters. The resolution is 0.15"/pixel and the image is
computed using star particles in the redshift range $0.1<z<5.8$.
Dust extinction and non-stellar sources are not taken into account
in this mock image.}} \label{fig:lightcone}
\end{center}
\end{minipage}
\end{figure*}

While in low-mass galaxies supernovae can deplete cold-gas
reservoirs and regulate star formation \citep[e.g.][]{Dekel1986},
they are ineffective in the deeper gravitational potential wells
of massive galaxies. In this regime, AGN offer a plausible source
of regulation, since the energy released by black-hole (BH)
accretion can be orders of magnitude larger than the binding
energy of the gas \citep[e.g.][]{Fabian2012}. However,
observational evidence for this process remains mixed.
{\color{black}While there is strong evidence that AGN regulate
cooling from hot gas \citep[e.g.][]{Tabor1993,McNamara2007}, the
AGN couple only weakly to gas acquired via mergers and accretion,
at least at low redshift \citep[e.g.][]{Kaviraj2015,Sarzi2016}}.
Nevertheless, both theoretical and observational work indicates
that AGN are likely to play an important role in shaping galaxy
evolution
\citep[e.g.][]{Silk1998,Bower2006,Croton2006,Cattaneo2009,Kimm2012,Dubois2012}.

While the interplay between star formation and feedback processes
shapes stellar-mass and black-hole growth, the observed changes in
the morphological mix of the Universe
\citep[e.g.][]{Mortlock2013,Conselice2014} is postulated to be
driven largely by merging. Both major mergers
\citep[e.g.][]{Toomre1972,White1978b,Barnes1992,Hernquist1992,Springel2005}
and minor interactions (which can trigger disc instabilities, see
e.g. Dekel et al. 2009; Kaviraj et al. 2013; Welker et al. 2015a),
can convert rotation-dominated (disky) systems into
dispersion-dominated spheroids.

Over the last two decades, semi-analytical models
\citep[e.g.][]{White1978a,White1991,Kauffmann1993,Hatton2003,Baugh2006,Benson2012}
have been successful in reproducing many of the bulk properties of
galaxies over a significant fraction of cosmic time.
{\color{black}By employing approximations derived from more
detailed numerical simulations, and empirical calibrations from
data, the semi-analytical approach has offered a computationally
inexpensive route to probing the phenomenology of galaxy formation
and, in particular, the theoretical analysis of today's large
survey datasets
\citep[e.g.][]{Cole2000,Hatton2003,Bower2006,Somerville2012,Overzier2013,Bernyk2016}.
A significant recent advance has been the advent of full
hydro-dynamical simulations in cosmological volumes which, while
more computationally demanding than the semi-analytical approach,
evolve the dark matter and baryons self-consistently. These
simulations provide predictions for the gas and baryonic
components of galaxies at high spatial resolution (on $\sim$kpc
scales) and, while sub-grid prescriptions are still needed, these
are applied on $\sim$kpc scales rather than at $\sim100$ kpc
scales as in the semi-analytical approach.}

The cosmological box sizes of the current generation of
hydro-dynamical simulations offer, for the first time, detailed
survey-scale predictions that can be compared to contemporary
datasets across a large fraction of cosmic time
\citep[e.g.][]{Devriendt2010,Schaye2010,Dubois2014,Vogelsberger2014,Schaye2015,Khandai2015}.
In concert with current and forthcoming observational data, these
simulations will, over the next few years, play a central role in
advancing our understanding of the key processes that drive galaxy
evolution, particularly in the (still poorly understood)
high-redshift Universe. It is, therefore, important to study the
reproduction of galaxy properties in such models to establish
whether they provide a reliable framework for interpreting current
and future observational datasets.

In this paper, we explore the predicted redshift evolution of
galaxy properties in the Horizon-AGN cosmological simulation
\citep{Dubois2014}, the first such simulation that uses an
adaptive mesh refinement (AMR) code and reaches $z=0$. Recent work
has used this simulation to probe the intrinsic alignments of
galaxies for calibrating weak-lensing analyses (Dubois et al.
2014; Welker et al. 2015a,b; Codis et al. 2015; Chisari et al.
2016), the evolution of galaxy morphology (Welker et al. 2015a)
and the role of merging at high redshift \citep{Kaviraj2015}. This
study is part of a series of papers that will explore the cosmic
evolution of black holes (Volonteri et al. 2016, Beckmann et al.
in prep), the morphological transformation of galaxies (Dubois et
al. in prep) and dark-matter cusp-core modulation by AGN activity
(Peirani et al. in prep).

In this work, we study the predicted evolution, in Horizon-AGN, of
quantities that are sensitive to the aggregate star formation
history of the galaxy population: luminosity and stellar mass
functions, the star formation main sequence, rest-frame
UV-optical-near infrared colours and the cosmic star formation
history. By comparing these predictions to an array of
corresponding observational data in the redshift range $0<z<6$, we
explore how well the simulation captures the evolutionary trends
of observed galaxies, and probe its usefulness as a tool to
investigate the processes that drive that evolution.

This paper is organized as follows. In Section 2, we describe the
simulation and the methodology used for the prediction of galaxy
luminosities, stellar masses and rest-frame colours. In Section 3,
we study the reproduction of luminosity functions, the star
formation main sequence and rest-frame UV-optical-near infrared
colours in Horizon-AGN. In Section 4, we probe the reproduction of
stellar-mass functions and the cosmic star formation history in
the model. We summarize our findings in Section 5.


\section{The Horizon-AGN simulation}
Horizon-AGN is a cosmological hydro-dynamical simulation
\citep{Dubois2014} that employs the adaptive mesh refinement
Eulerian hydrodynamics code, RAMSES \citep{Teyssier2002}. While
the simulation is described in \citet{Dubois2012} and
\citet{Dubois2014}, we briefly revisit key aspects of the model
here. The size of the simulation box is 100 $h^{-1}\, \rm Mpc$
(comoving), which contains $1024^3$ dark matter particles and uses
initial conditions corresponding to a \emph{WMAP7} $\Lambda$CDM
cosmology \citep{Komatsu2011}. The dark matter mass resolution is
8 $\times$ 10$^7$ M$_{\odot}$. A quasi Lagrangian criterion is
used to refine the initially uniform $1024^3$ grid, when 8 times
the initial total matter resolution is reached in a cell, down to
a minimum cell size of $1 \, \rm kpc$ in proper units. Gas cools
via H, He and metals (following Sutherland \& Dopita 1993) down to
10$^4$ K, and a uniform UV background is switched on at $z = 10$,
following \citet{Haardt1996}.


\subsection{Star formation and stellar feedback}
Star particles are created using a standard 2\% efficiency per
free fall time \citep{Kennicutt1998}, when the gas hydrogen
density reaches a critical threshold of 0.1 H cm$^{-3}$. Star
formation is assumed to follow a Schmidt-Kennicutt law
\citep{Kennicutt1998}, with a Poissonian random process
\citep{Rasera2006,Dubois2008} that has a stellar mass resolution
of $\sim$2 $\times$ 10$^6$ M$_{\odot}$.

We implement a subgrid model for stellar feedback that probes all
processes that may impart thermal and kinetic feedback on the
ambient gas. Many previous works that implement stellar feedback
employ a single supernova explosion per star particle to minimize
computational cost \citep[e.g.][]{Dubois2008}. However, this is an
oversimplification, particularly from the point of view of
chemical enrichment. A significant fraction of stellar mass is, in
fact, lost through various phases of stellar evolution, such as
Wolf-Rayet stars or the asymptotic giant branch
\citep{Leitherer1992}. Thus, stellar feedback should be modelled
more realistically by taking into account stellar winds and both
Type II and Type Ia supernovae \citep[SNe;
e.g.][]{Kobayashi2011,Hopkins2012}.

To this end, Horizon-AGN implements continuous stellar feedback
that includes momentum, mechanical energy and metals from Type II
SNe, stellar winds, and Type Ia SNe. For stellar winds and Type II
SNe, \texttt{Starburst99} \citep{Leitherer1999,Leitherer2010} is
used to generate look-up tables as a function of metallicity and
age. Specifically, we use the Padova model \citep{Girardi2000}
with thermally pulsating asymptotic branch stars
\citep{Vassiliadis1993}, with the kinetic energy of stellar winds
calculated via the `Evolution' model of \citet{Leitherer1992}.

We implement Type Ia SNe following \citet{Matteucci1986}, assuming
a binary fraction of 5\% \citep{Matteucci2001}. The chemical
yields for Type Ia explosions are taken from the W7 model of
\citet{Nomoto2007}. Although the energy input from this source is
minor compared to that of Type II SNe ($\approx 10\%$ of the total
kinetic energy), they provide a significant fraction ($\approx
50\%$) of the iron for the chosen parameters. In order to mimic
the propagation of bubbles as realistically as possible, we allow
for the injection of energy, mass and momentum only if a blast
wave from star particles in each cell propagates to $r_{\rm B} \ge
2\,\Delta\, x$, where $\Delta\, x$ is the size of the host cell
and $r_{\rm B}$ is the radius of the shock front at $\Delta t$:

{\color{black}\begin{equation} r_{\rm B} \approx 44\,{\rm pc}\,
\left[ \left( \frac{E}{10^{47}\,{\rm erg}}\right)
\left(\frac{0.1\,{\rm H/cm^3}}{n_{\rm H}}\right)\right]^{0.2}
\left(\frac{\Delta t}{10^7\,{\rm yr}}\right)^{0.4}.
\end{equation}\\}

If the energy released from each cell at $\Delta t \equiv t_{\rm
last} - t_{\rm now}$ is not large enough to push the blast wave to
$2\,\Delta\, x$,  we accumulate the energy, momentum, and metals
until the next time step, where $t_{\rm last}$ is the time at
which the last blast wave is launched. This produces a more
realistic evolution of bubbles and prevents them expanding too
rapidly.

To reduce computational cost, we model the stellar feedback as a
heat source after 50 Myr, while the energy liberated before 50 Myr
from star particles is modelled as kinetic feedback, as described
above. This is a reasonable choice, given that, after 50 Myr,
almost all of the energy is liberated via Type Ia SNe that have
time delays between several hundred Myrs to a few Gyrs
\citep[e.g.][]{Maoz2012}. These systems are less prone to
excessive radiative losses, as stars are likely to disrupt or move
away from their dense birth clouds after around a few tens of Myrs
\citep[e.g.][]{Blitz1980,Hartmann2001}.


\subsection{Feedback from black holes}
Seed black holes (BHs) with a mass of $10^5\,\rm M_{\odot}$ are
assumed to form in dense star forming regions where both the gas
and stellar densities are above $\rho_0$, and where the stellar
velocity dispersion is larger than $100 \,\rm km\,s^{-1}$. The
growth of the BH is tracked self consistently, based on a modified
Bondi accretion rate at high gas densities \citep{Booth2009}. The
accretion rate is capped at Eddington, with a standard radiative
efficiency of $0.1$.

The central BH impacts ambient gas in two possible ways, depending
on the gas accretion rate. For Eddington ratios $>0.01$ (high
accretion rates), 1.5\% of the accretion energy is injected as
thermal energy (a quasar-like feedback mode), whilst for Eddington
ratios $<0.01$ (low accretion rates), bipolar jets are employed
with a 10\% efficiency. The parameters are chosen to produce
agreement with the local cosmic black-hole mass density, and the
M$_{\rm BH}$ - M$_*$ and M$_{\rm BH}$ - $\sigma_*$ relations
\citep{Dubois2012}. An explicit dynamical drag force is exerted
from the gas onto the BHs \citep{Ostriker1999,Chapon2013} in order
to stabilize BH motions into galaxies and suppress limited
resolution effects \citep{Dubois2013}. Finally, BHs are allowed to
merge when they are closer than $4 \,\rm kpc$ and when their
relative velocity is smaller than the escape velocity of the
binary.

We note that, apart from choosing the BH-feedback parameters to
match the M$_{\rm BH}$ - M$_*$ and M$_{\rm BH}$ - $\sigma_*$
relations at $z=0$, Horizon-AGN is not otherwise tuned to
reproduce the bulk properties of galaxies such as stellar mass and
luminosity functions, galaxy sizes etc. at $z\sim0$.


\subsection{Galaxy identification and prediction of observables}
We identify galaxies using the \texttt{AdaptaHOP} structure finder
\citep{Aubert2004,Tweed2009}, applied to the distribution of star
particles. Structures are selected using a local threshold of 178
times the average matter density, with the local density of
individual particles calculated using the 20 nearest neighbours.
Only structures that have more than 50 particles are considered.
We compute galaxy fluxes and magnitudes using the stellar models
of \citet[][BC03 hereafter]{BC2003}, using a \citet{Chabrier2003}
initial mass function (IMF). We assume that each star particle
behaves as a simple stellar population (SSP), and compute its
contribution to the total spectral energy distribution (SED) by
logarithmically interpolating the models in metallicity and age
and multiplying by the initial mass of the particle.
{\color{black}Figure 1 shows a 14 arcmin$^2$ mock $u,r,z$
composite image from the Horizon-AGN lightcone (a simulated box
where one axis is redshift), constructed using star particles in
the redshift range $0.1 < z < 5.8$, via the BC03 models as
described above. The resolution of the image is 0.15"/pixel and
the lightcone was produced on-the-fly at every time step of the
simulation. We direct readers to \citet{Pichon2010} for details of
the lightcone construction. Note that dust extinction is not taken
into account and that non-stellar sources (e.g. AGN) are not
included in this image.}

We calculate attenuation by dust using the \texttt{SUNSET} code.
The gas density and metallicity are extracted, under the
assumption that the dust mass scales with the gas metal mass, with
a dust-to-metal ratio of 0.4 \citep[e.g.][]{Dwek1998,Draine2007}.
We compute the column density of dust, and thus the line-of-sight
optical depth for each star particle, using the $R=3.1$ Milky Way
dust grain model of \citet{Weingartner2001}. The gas is assumed to
be transparent beyond 1 virial radius (but we note that relaxing
this assumption leaves our conclusions unchanged). This dust
implementation assumes that all the dust is placed in a screen in
front of each star particle. Nevertheless, the geometry of the
metals, and therefore the spatial distribution of dust within the
galaxy, is taken into account. The total dust-attenuated SED is
computed by summing the contribution of all star particles.
Magnitudes are computed after convolving this SED through the
appropriate filtercurves. Galaxy magnitudes are extracted within
Petrosian apertures \citep{Petrosian1976,Blanton2001,Graham2006}.
While many observational studies use Petrosian apertures, other
variants, such as Kron \citep[e.g.][]{Kron1980} or modified Sersic
\citep[e.g.][]{Bernardi2013} apertures can also be employed,
leading to small shifts in galaxy luminosities
\citep[e.g.][]{Graham2005}. However, these offsets are typically
lower than the statistical uncertainties.

We note that, ideally we would like to quantify the effect of dust
on galaxy SEDs via a full radiative transfer approach. While
processing all model galaxies using this approach is prohibitively
time-consuming, we check, in Section 3 below, whether using a dust
screen is a reasonable assumption to make in our calculations. We
perform this check by using \texttt{SUNRISE}
\citep{Jonsson2006,Jonsson2010}, a Monte-Carlo radiative transfer
code applicable to arbitrary dust geometries, on a small random
sample of galaxies. In a similar vein to \texttt{SUNSET},
\texttt{SUNRISE} employs the gas metallicity distribution as a
proxy for the dust distribution, with a metal-to-dust ratio of 0.4
and the $R=3.1$ Milky Way dust grain model of
\citet{Weingartner2001} to set the dust grain composition and
size. A sub-resolution model for photo-dissociation regions
(\texttt{MAPPINGSIII}; Dopita et al. 2005, Groves et al. 2008) is
used to account for obscuration of young stars by their birth
clouds. \texttt{SUNRISE} then performs radiative transfer ray
tracing to calculate the dust absorption and scattering. The
attenuated SED is convolved with the relevant filtercurves to
extract rest-frame magnitudes.

As shown below, good agreement is found between the predicted
magnitudes from the screen (\texttt{SUNSET}) and radiative
transfer (\texttt{SUNRISE}) approaches in the optical filters,
with small offsets in the UV wavelengths. Overall, the dust screen
assumption appears sufficient for the purposes of this study.
However, as we note in the following sections, the uncertainties
in the modelling of dust are large and are, in particular,
difficult to quantify for high-redshift galaxies without future
datasets. The comparison of rest-frame UV-optical-near infrared
colours, while useful, may not offer the best test of the overall
performance of the model.



\begin{figure*}
\begin{minipage}{172mm}
\begin{center}
$\begin{array}{cc}
\includegraphics[width=3.1in]{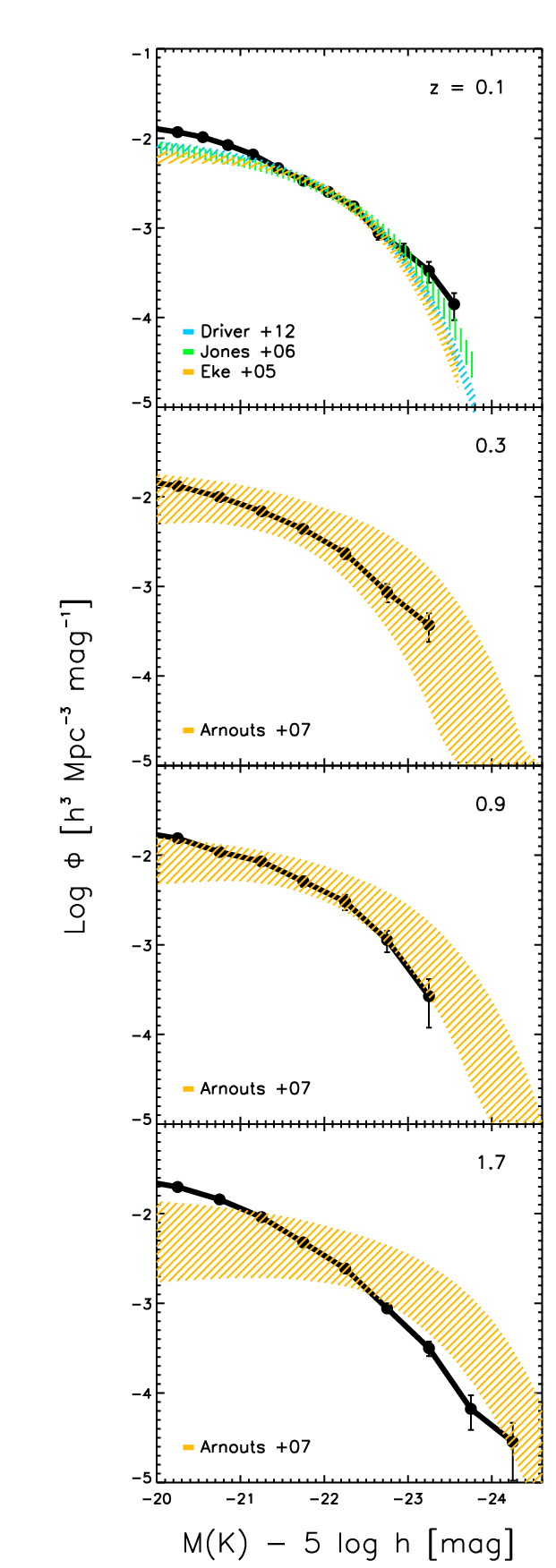} & \includegraphics[width=3.1in]{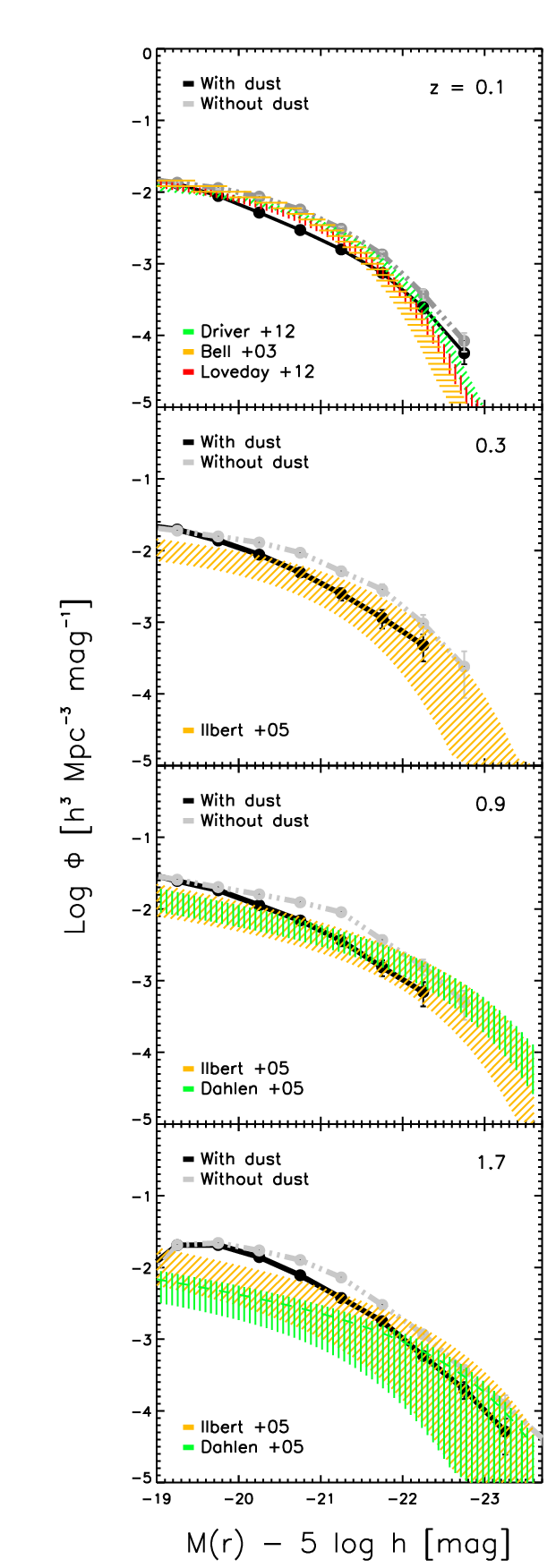}
\end{array}$
\caption{Comparison of the predicted $K$ (left-hand column) and
$r$-band (right-hand column) luminosity functions from Horizon-AGN
to observational data. The solid lines show dust-attenuated
luminosity functions predicted by Horizon-AGN, while the
grey-dotted curves show their unattenuated counterparts. Note
that, given its low sensitivity to dust, the rest-frame $K$-band
luminosity function with and without extinction are almost
identical - we only show the dust-reddened $K$-band luminosities
here. Observational data is shown using the coloured hatched
regions (see legend for the individual datasets used).}
\label{fig:lf}
\end{center}
\end{minipage}
\end{figure*}


\begin{figure*}
\begin{minipage}{172mm}
\begin{center}
$\begin{array}{c}
\includegraphics[width=0.92\textwidth]{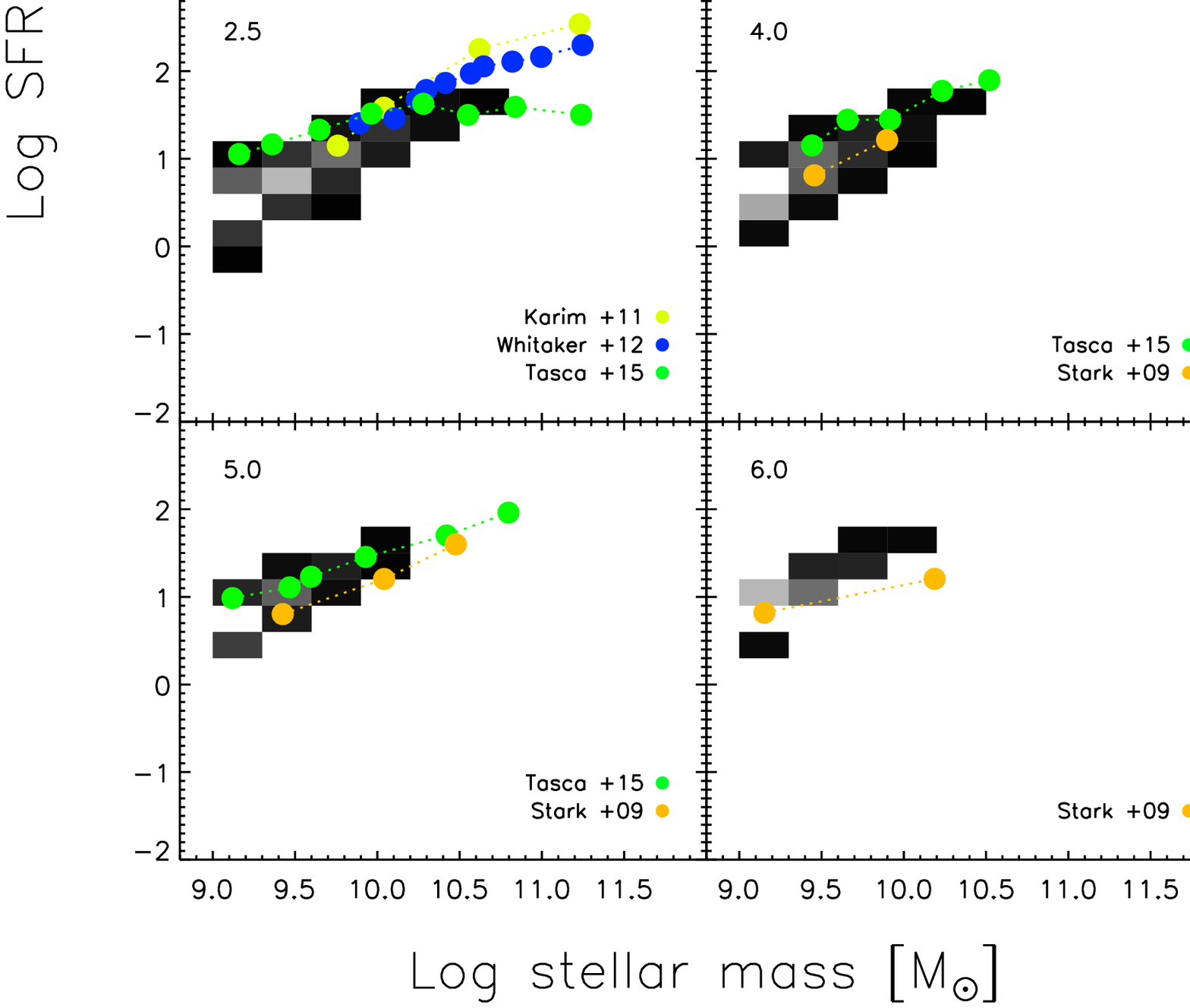}
\end{array}$
\caption{Comparison of the predicted star formation main sequence
in Horizon-AGN to observations in the redshift range $0<z<6$.
Observational data is taken from \citet{Karim2011}, who estimate
SFRs via 1.4 GHz fluxes, and \citet{Lee2015},
\citet{Whitaker2012}, \citet{Tasca2015} and \citet{Stark2009}, who
estimate SFRs via multi-wavelength SED fitting. The Hess diagram
indicates the predicted star formation main sequence from
Horizon-AGN (darker shades indicating lower galaxy density). } \label{fig:sfr}
\end{center}
\end{minipage}
\end{figure*}


\begin{figure*}
\begin{minipage}{172mm}
\begin{center}
$\begin{array}{c}
\includegraphics[width=0.92\textwidth]{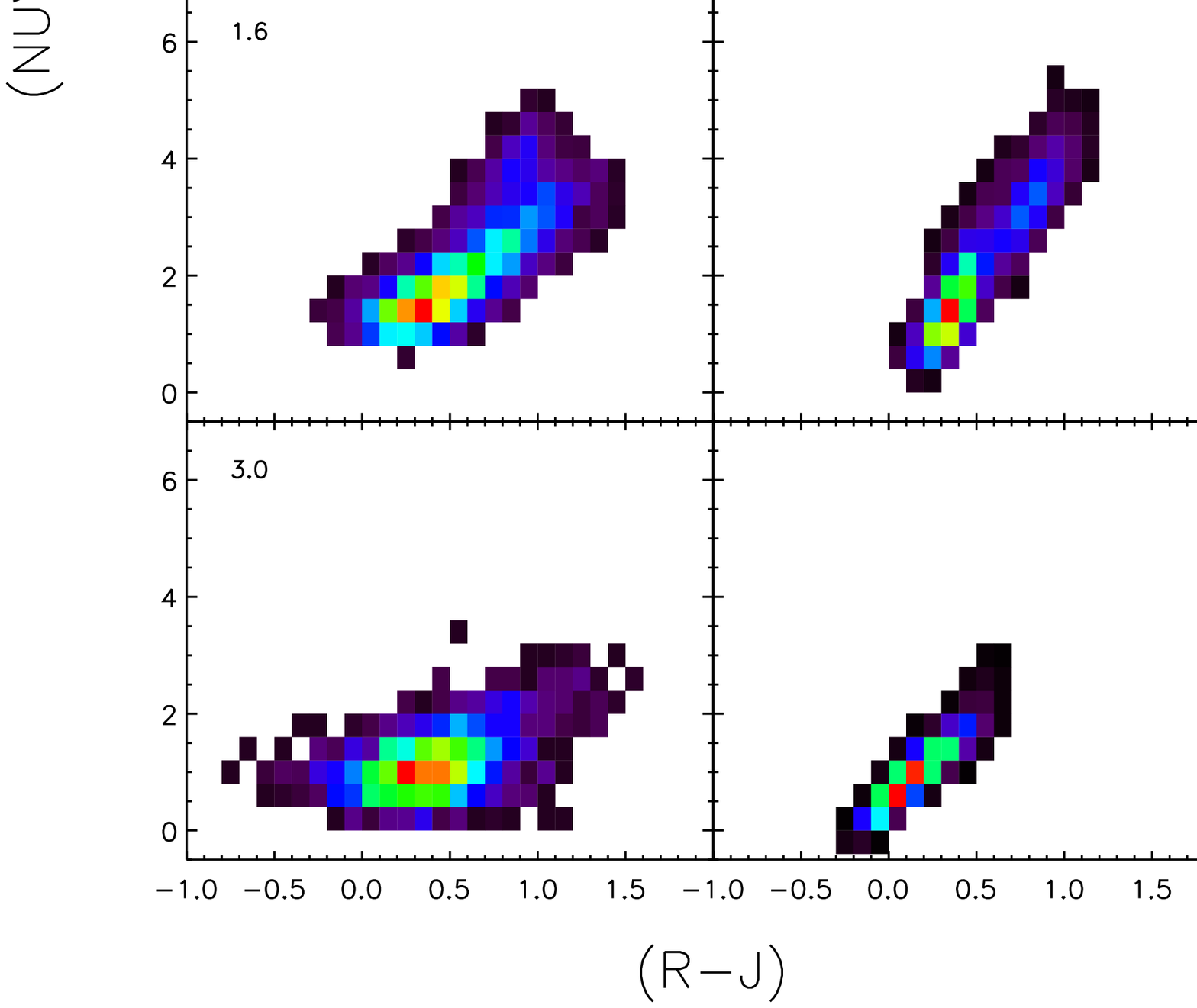}
\end{array}$
\caption{Comparison of the predicted evolution of rest-frame
$NUV-R-J$ colours in Horizon-AGN to observational data from the
COSMOS2015 catalogue (Laigle et al. submitted). The left-hand
column shows observational data, while the right-hand column shows
the predictions from Horizon-AGN. The Hess diagrams indicate the
galaxy density (red = highest density, black = lowest density).}
\label{fig:cc}
\end{center}
\end{minipage}
\end{figure*}


\section{Luminosities, star formation rates and colours}
We begin by comparing the predictions of Horizon-AGN to observed
luminosity functions, since these are one of the basic quantities
delivered by observations. Since the ages and metallicities of
star particles are known precisely in the simulation, galaxy
luminosities can be constructed, the main uncertainty in the
predicted luminosity being the derivation of the dust mass from
the gas and metal content of the model galaxy
\citep[e.g.][]{Guiderdoni1987,Devriendt1999}. We note that the
simulated gas-phase metallicity in Horizon-AGN is under-estimated
by a factor of $\sim$ 2 to 4 compared to observations
\citep[e.g.][]{Maiolino2008,Mannucci2009}. This happens because
our blast wave model allows for the propagation of energy and
metals only when it reaches $2\Delta x$, where $\Delta x$ is the
size of a host cell of SNe. Given the relatively low resolution
adopted in Horizon-AGN, we find that this tends to delay the metal
enrichment of star-forming clouds in the simulation, particularly
when the specific star formation rate is low (i.e. at lower
redshift).

{\color{black}To correct for these lower metallicities, we
calibrate the gas-phase metallicities by multiplying a
redshift-dependent renormalisation factor ($f_{\rm no}$) that
brings the simulated metallicity in agreement with the observed
mass-gas phase metallicity relations at $z=0$, 0.7, 2.5 and 3.5
\citep{Maiolino2008,Mannucci2009}, where $f_{\rm no}=4.08 - 0.21z
- 0.11z^2$. These authors calculate their mass-metallicity
relations using strong line diagnostics (e.g. [OIII]5007/H$\beta$,
[OIII]5007/[OII]3727, [NeIII]3870/[OII]3727), using the direct
$T_e$ method for low metallicities (12+log(O/H)$<$8.35) and the
photoionisation models of \citet{Kewley2002} for higher
metallicities (which are similar to the ones from
\citet{Kewley2008}). We do not attempt to make the calibration
stellar mass-dependent, but simply adjust the normalization as a
function of redshift, following $f_{\rm no}$. This is because the
shape of the predicted mass-gas phase metallicity relation is
reasonably consistent with the shape of their observed
counterparts at redshifts where data is available, although it
should be noted that the data do not extend across the entire mass
range spanned by the model galaxies. In Appendix A we show the
mass-gas phase metallicity relations predicted by the simulation,
the observational data to which we perform the calibrations and
the corrected relations that are used in calculating the
properties of model galaxies in the simulation.}

In Figure \ref{fig:lf}, we compare the predicted rest-frame $K$
and $r$-band luminosity functions in Horizon-AGN to observational
data in the redshift range $0<z<2$. Observational estimates of
rest-frame $K$-band luminosities are largely restricted to this
redshift range because observational data currently extends into
the observed mid-infrared, from facilities like \emph{Herschel}
\citep[e.g.][]{Pilbratt2010}. Rest-frame $K$ corresponds to the
longest wavelengths at which stellar light still dominates the
SED, the inter-stellar medium becoming increasingly important
longward of this filter \citep[e.g.][]{Fazio2004,Eales2010}. In
addition, its negligible sensitivity to dust and young stars makes
the $K$ band a good tracer of the underlying stellar mass of the
galaxy \citep[e.g.][]{Kauffmann1998}\footnote{It is worth noting
that intermediate-age populations, like the thermally-pulsating
AGB (TP-AGB), may be bright in the near-infrared wavelengths
\citep[e.g.][]{Maraston2006}. However, while at low redshift the
contribution of these populations will be minimal due to the low
star formation rate, their contribution is likely to continue to
be negligible across the redshift range probed by our
luminosity-function analysis \citep{Kriek2010,Zibetti2013}.}. The
$r$-band, being a shorter wavelength filter, is more sensitive to
the mass-to-light ratio of the galaxy, which in turn depends on
its star formation history.

In Figure \ref{fig:lf}, we compare the predicted luminosity
functions with and without dust extinction to their observed
counterparts. Given its low sensitivity to dust, the rest-frame
$K$-band luminosity function with and without extinction are
almost identical. Thus, we only show the dust-reddened $K$-band
luminosities. The shaded regions indicate the uncertainties in the
observed luminosity functions, based on the errors in the fitted
Schechter-function \citep{Schechter1976} parameters (but not
including the effect of cosmic variance). {\color{black}In our
redshift range of interest, the slope and normalisation of the
predicted $K$-band luminosity function from Horizon-AGN agrees
reasonably well with its observational counterparts, within the
observational uncertainties. However, some points of tension with
the data are worth noting. In the local Universe ($z\sim0.1$), the
simulation does not reproduce all observational datasets equally
well, overshooting the \cite{Eke2005} data at both the low and
high luminosity ends. While the predicted data points generally
fall within the observed ranges, the predicted luminosity
functions are slightly steeper than their observed counterparts at
all redshifts. The difference in steepness is most apparent at
$z\sim1.7$ where it causes the predictions to overshoot the data
at high luminosities and undershoot them at low luminosities.
Indeed, the tendency to overshoot at low luminosities mirrors a
trend seen in the mass-function analysis for low-mass galaxies,
and we return to this point in Section 4 below. Nevertheless, the
generally good reproduction of the evolving $K$-band luminosity
function, within the observational uncertainties, indicates that,
on average, the aggregate mass growth of galaxies over cosmic time
is well reproduced by the simulation (this is also borne out by
the mass function analysis presented later in this study). Similar
trends are seen in the comparison of the predictions to the
observed $r$-band luminosity functions. While the model
predictions agree with data within observational uncertainties,
the overproduction of galaxies at the low-luminosity end is also
apparent in the $r$-band filter.}


\begin{figure}
\begin{center}
$\begin{array}{c}
\includegraphics[width=3.3in]{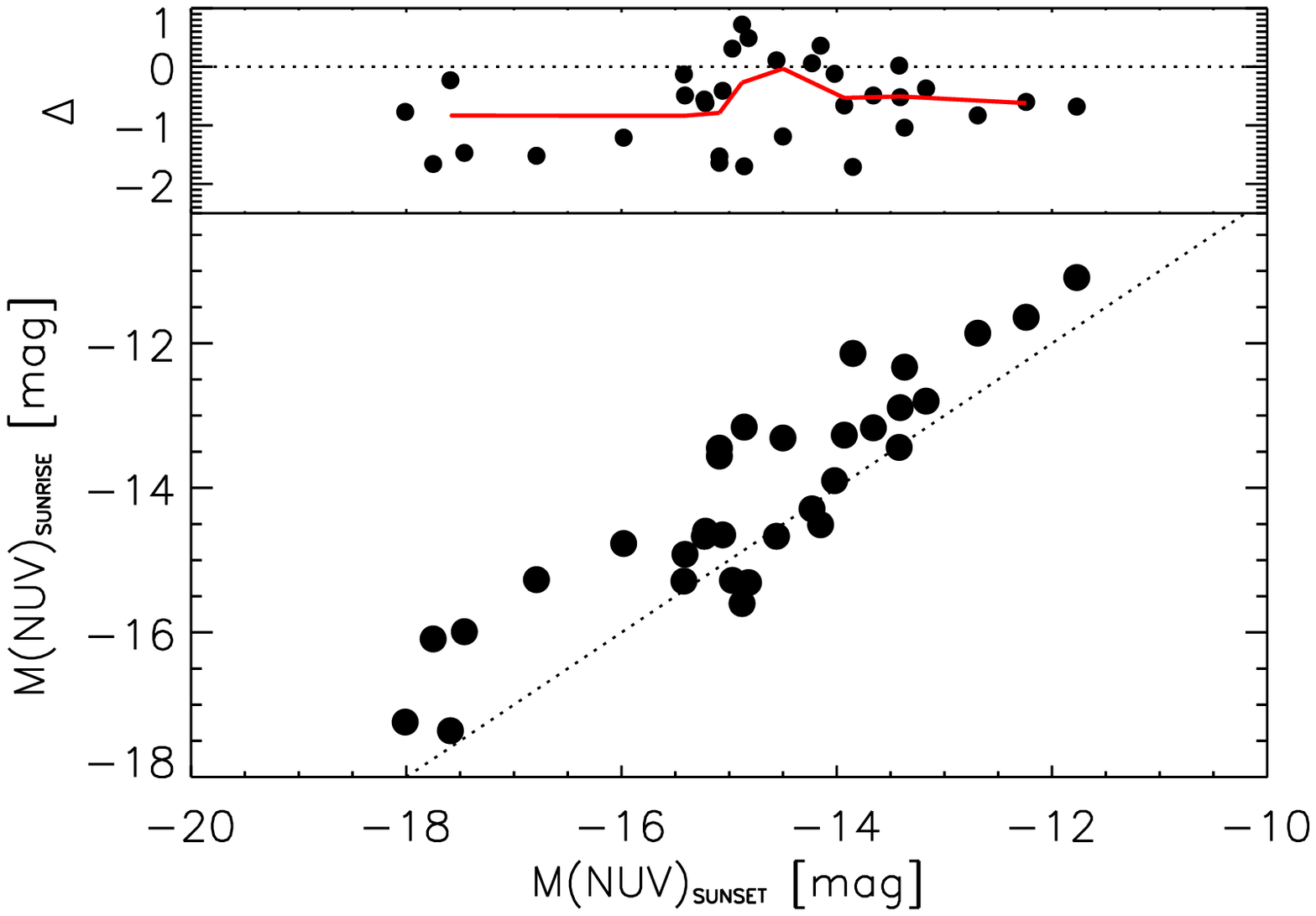}\\
\includegraphics[width=3.3in]{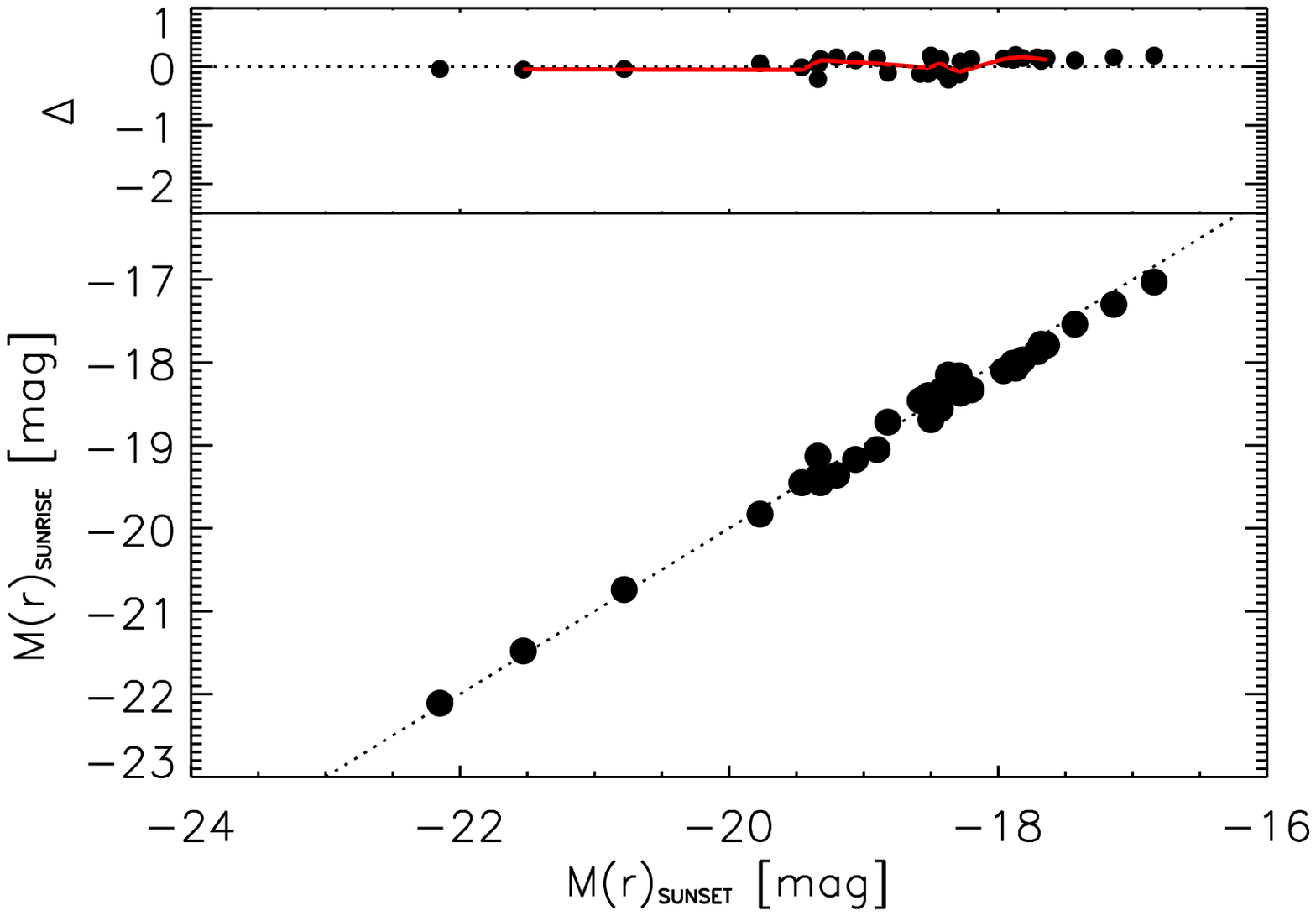}\\
\includegraphics[width=3.3in]{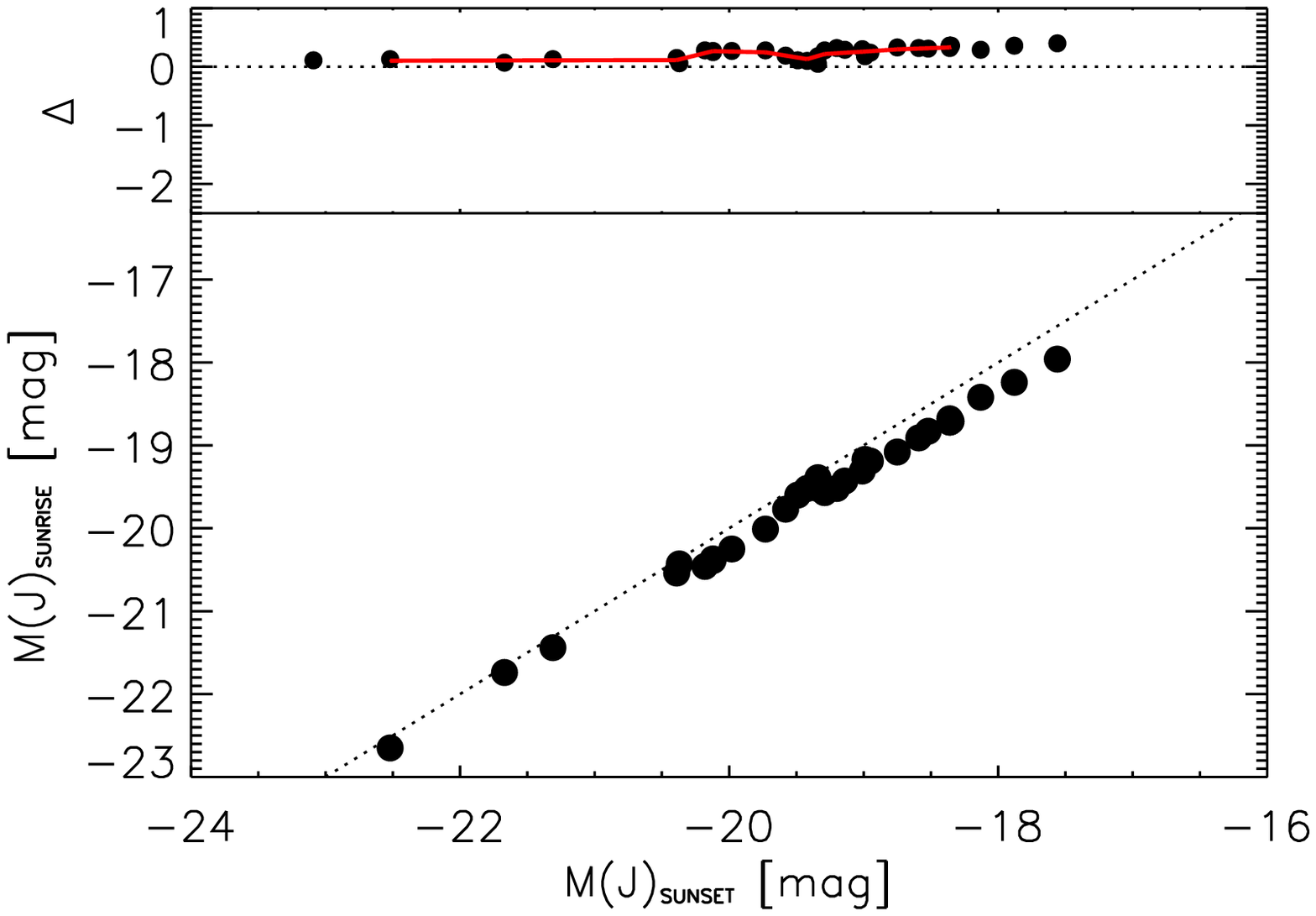}
\end{array}$
\caption{Comparison of the predicted magnitudes in the $NUV$
(top), $r$ (middle) and $J$ (bottom) band filters, using
\texttt{SUNRISE} (full radiative transfer) and \texttt{SUNSET}
(dust screen in front of each star particle), on a small random
sample of Horizon-AGN galaxies at $z\sim0.1$. The offset in these
predicted magnitudes ($\Delta$) is shown in the top section of
each panel, with $\Delta \equiv$
M$_{\textnormal{\texttt{SUNRISE}}}$ -
M$_{\textnormal{\texttt{SUNSET}}}$.} \label{fig:sunrise_sunset}
\end{center}
\end{figure}


We proceed by comparing the `star-formation main sequence' (i.e.
the star formation rate plotted against stellar mass) predicted by
Horizon-AGN to observational data in the redshift range $0<z<6$
(Figure \ref{fig:sfr}). Comparison to the observed main sequence
probes whether the instantaneous star-formation activity in the
simulation is consistent with observations. Performing this
exercise over a large redshift range then offers insights into how
well Horizon-AGN predicts stellar mass growth over cosmic time,
compared to what is observed in the real galaxy population. Figure
\ref{fig:sfr} indicates that, within the dispersion between the
observed main sequences, the simulation produces good agreement
with the observations, across our redshift range of interest
($0<z<6$). {\color{black}The simulated main sequence appears to
fall slightly below the observed ones at $z\sim1.7$ and the
predictions are inconsistent, at this redshift, with the Karim et
al. and Whitaker et al. datasets at high stellar masses, while
matching the Tasca et al. data well.} However, we note that the
observed loci shown are median values and that the observed main
sequences typically have spreads of $\sim$0.5 dex
\cite[e.g.][]{Whitaker2012}, suggesting that the overlap between
the theoretical and observational values is also reasonable at
this redshift. Overall, our results indicate that Horizon-AGN
predicts the observed main sequence with good accuracy between
$z=0$ and $z=6$, suggesting that the aggregate stellar mass growth
of galaxies with stellar masses greater than $\sim$10$^9$
M$_{\odot}$ is generally well reproduced by the simulation.

We next compare the predicted evolution of rest-frame colours in
Horizon-AGN to observational data. To perform this exercise in a
consistent way across redshift, we use observational data from a
new homogeneous multi-wavelength catalogue of a single area of
sky: COSMOS2015 (Laigle et al. submitted). The COSMOS2015 catalog
contains 30 band-photometry, from UV to IR (0.25-7.7$ \mu$m) of
more than half a million objects in the 2 deg$^2$ COSMOS field. It
includes the optical datasets from previous releases
\citep{Capak2007,Ilbert2009}, the new $Y$-band data taken with the
Subaru Hyper-Suprime-Cam (PI: Guenther), new near-infrared (NIR)
data from the UltraVISTA-DR2 survey and IR data from the SPLASH
program (P. L. Capak et al. in prep.). {\color{black}The
photometry for the COSMOS2015 catalog} is extracted using
SExtractor \citep{Bertin1996} in dual image mode. The detection
image is the chi-squared sum of four NIR images from
UltraVISTA-DR2 ($Y,J,H,Ks$) and the optical $z^{++}$ band image
from Subaru SuprimeCam. Photometric redshifts, rest-frame
magnitudes and stellar masses are computed using SED fitting via
the \texttt{LePhare} code
\citep{Arnouts1999,Arnouts2002,Ilbert2006}, following the
methodology described in \cite{Ilbert2013}. Comparison to
published spectroscopic redshifts of galaxies in the COSMOS field
indicates a precision of 0.007 and a catastrophic failure rate of
0.5\% for bright galaxies ($i_{+}<22.5$). The deepest region of
the COSMOS2015 catalogue reaches a 90\% completeness limit for
galaxies with masses greater than 10$^{10}$ M$_{\odot}$, out to
$z=4$. Note that, since COSMOS2015 is a deep field, it does not
contain many galaxies in the local Universe. Our lowest redshift
bin in this analysis is therefore $z=0.3$.

In Figure \ref{fig:cc}, we compare the predicted evolution of
rest-frame $NUV-R-J$ colours in Horizon-AGN to observed galaxies
in COSMOS2015. The UV ($NUV-R$) colour traces very recent star
formation (stars with ages $<0.5$ Gyr), with even residual
($<1$\%) mass fractions of young stars capable of driving galaxies
into the UV blue cloud (Kaviraj et al. 2007a,b). The optical
($R-J$) colour, on the other hand, traces stellar mass growth over
several Gyrs in the past (Kaviraj et al. 2007a). Taken together,
these colours probe the formation history of the galaxy population
over the last few Gyrs.

Figure \ref{fig:cc} indicates that the model galaxies occupy
similar parts of the $NUV-R-J$ colour space as their observed
counterparts. While the simulated galaxies agree well with the
observed optical colours {\color{black}(although note that the
model occupies a narrower locus at $z\geq1.6$)}, the predicted
bimodality in the UV colour is weaker than in observational data,
especially at low redshift. However, the good reproduction of the
star formation main sequence and the optical colours indicates
that this is largely due to small amounts of residual star
formation in the model galaxies (recall that even negligible mass
fractions of young stars can produce blue UV colours). While, this
residual star formation has little bearing on the bulk
stellar-mass growth of the galaxy population as a whole, the
weakness of the bimodality in the predicted UV colours suggests
that the feedback recipes employed by the model do not quench star
formation as completely as is the case in real galaxies at low
redshift.

As noted before, a large uncertainty in this exercise is the way
the attenuation by dust is estimated for simulated galaxies. In
particular, it is worth exploring the robustness of the dust
screen assumption in \texttt{SUNSET}, by comparing these results
to a full radiative transfer treatment using \texttt{SUNRISE}. In
Figure \ref{fig:sunrise_sunset} we compare magnitudes computed
using \texttt{SUNSET} to their counterparts using
\texttt{SUNRISE}. As noted above, processing all simulated
galaxies using this approach is prohibitively time-consuming.
Therefore, we explore the potential differences between these two
approaches for a small, random sample of galaxies at $z\sim0.3$
that spans our mass range of interest.

\begin{figure}
\begin{center}
$\begin{array}{c}
\includegraphics[width=3.3in]{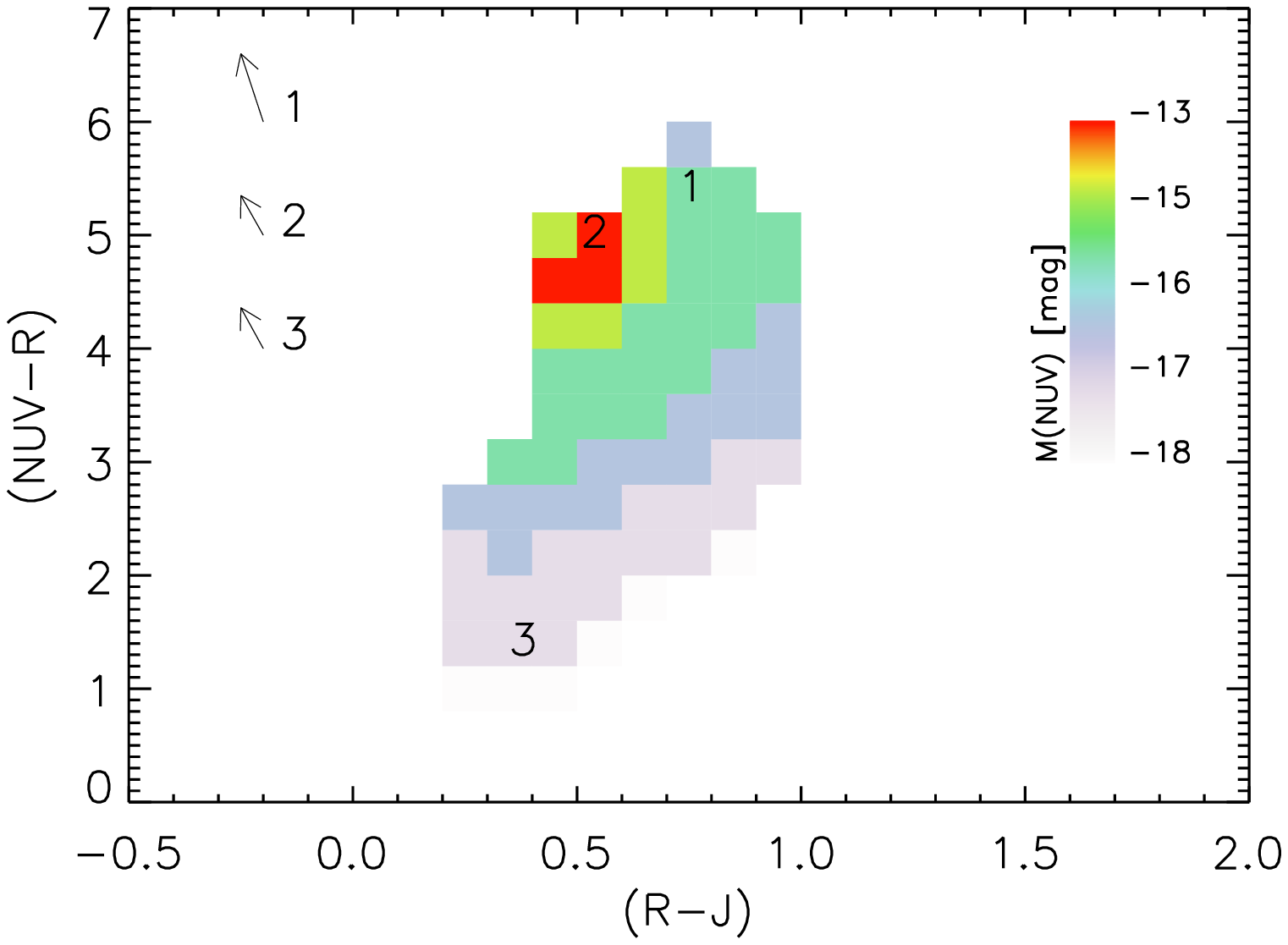}
\end{array}$
\caption{The predicted rest-frame $NUV-R-J$ colours in Horizon-AGN
at $z=0.3$, colour-coded by the absolute $NUV$ magnitude. The
arrows indicate how the predicted colours are likely to move if
the effect of dust were estimated using \texttt{SUNRISE} (full
radiative transfer) rather than \texttt{SUNSET} (dust screen in
front of each star particle).} \label{fig:sunrise_colour_offset}
\end{center}
\end{figure}

We find that, while the optical magnitudes are almost identical
using the two approaches, the predicted UV magnitudes are somewhat
fainter when they are calculated via full radiative transfer and
show a dependence on the UV magnitude itself (Figure
\ref{fig:sunrise_sunset}). The difference is due to the fact that
the optical depth due to both absorption and scattering is, on
average, larger in the shorter wavelengths. A simple screen
absorption model (as employed by \texttt{SUNSET}) is likely to
progressively underestimate the attenuation at increasingly
shorter wavelengths. In addition, our results suggest that the
contribution of scattering to the optical depth depends on the
dust geometry. Figure \ref{fig:sunrise_colour_offset} indicates
that applying the average values of the offsets in Figure
\ref{fig:sunrise_sunset} enhances the bimodality slightly, but not
enough to achieve reasonable agreement with the observational
data. Our conclusions above therefore remain unchanged - the
feedback recipes in the model appear unable to quench star
formation to the extent that is required to produce the bimodality
in rest-frame UV-optical-near infrared colours.


It is worth noting here that there are additional uncertainties in
the treatment of dust that may further complicate the comparison
between theory and observation. For example, while the
dust-to-metal ratio is assumed to be a fixed (Milky-Way-like)
value in our analysis, this value may not be applicable to all
galaxies. Studies of damped Lyman-$\alpha$ absorbers
\citep[e.g.][]{Vladilo2004} and the UV/optical afterglow spectra
of gamma ray burst host-galaxies at high redshift
\citep[e.g.][]{DeCia2013} indicate that dust-to-metal ratios may
vary with both metallicity and the metal column density \citep[see
also][]{Fisher2014,Herrera-Camus2012}. In a similar vein, the
extinction law (assumed in this study to be Milky-Way-like), may
also vary as a function of galaxy properties like age and
metallicity \citep[e.g.][]{Buat2012}. The potentially large
unknowns in the treatment of dust makes the prediction of colours,
especially those involving shorter wavelengths, uncertain. In
particular, the current dearth of observational data at high
redshift makes it difficult to ascertain whether the properties of
dust in our local neighbourhood can be blindly extrapolated to
early epochs. In that sense, the rest-frame UV-optical-near
infrared colour space may not be the best test of the reliability
of the model.


\begin{figure*}
\begin{minipage}{172mm}
\begin{center}
$\begin{array}{c}
\includegraphics[width=0.90\textwidth]{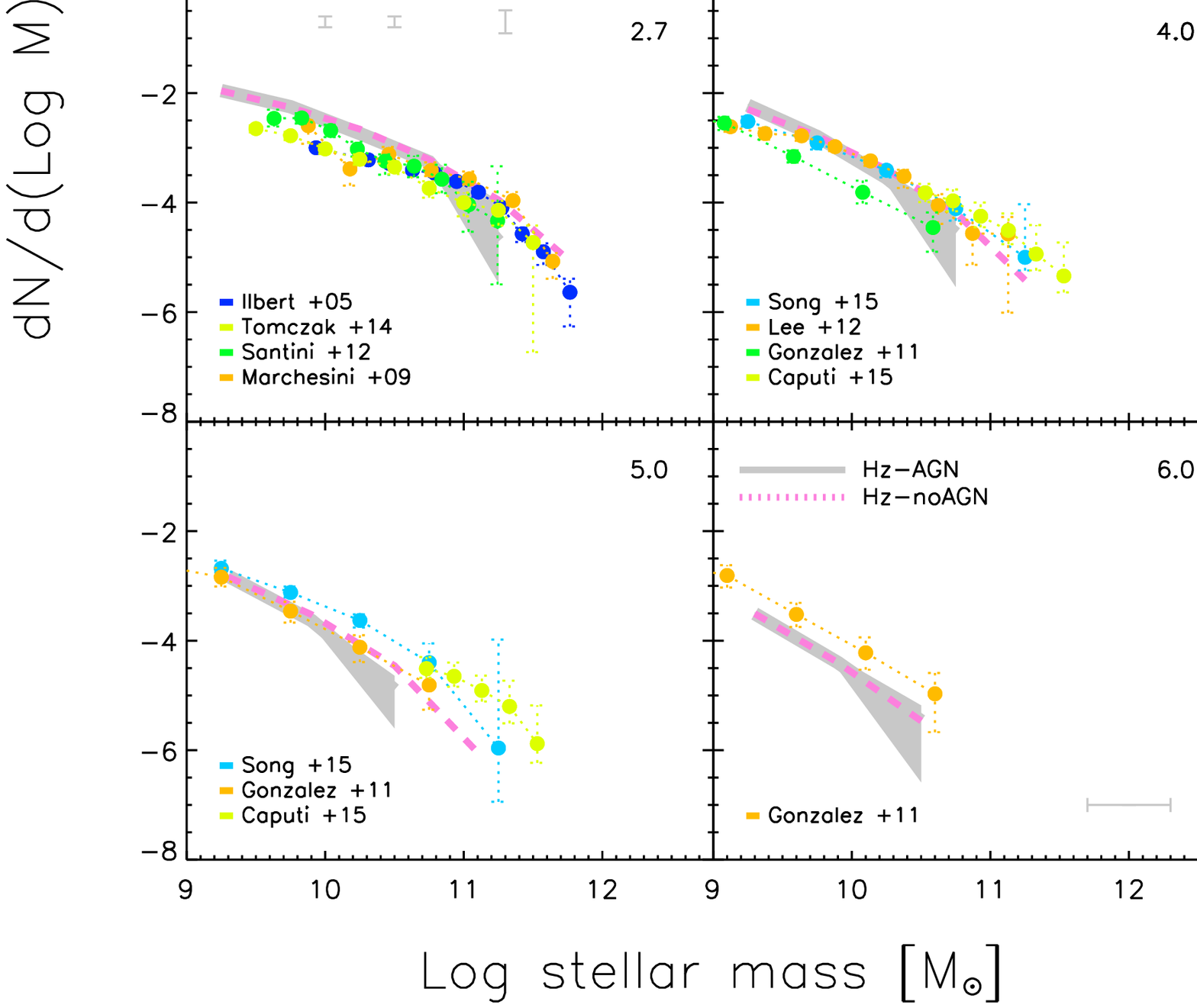}
\end{array}$
\caption{Comparison of the predicted stellar mass function to
observational data in the redshift range $0<z<6$. The grey shaded
region shows the prediction from Horizon-AGN (with the width of
the region indicating Poisson uncertainties). The pink dashed
curves indicate predictions from Horizon-noAGN, a twin simulation
without BH feedback. Vertical error bars indicate observational
uncertainties due to cosmic variance \citep{Ilbert2013}. The
horizonal error bar (0.3 dex) indicates typical observational
uncertainties in stellar masses derived from SED fitting
\citep[e.g.][]{Muzzin2009}.} \label{fig:mf}
\end{center}
\end{minipage}
\end{figure*}


Nothwithstanding the dust-related uncertainties outlined above,
our analysis indicates that Horizon-AGN produces good agreement
with the redshift-evolution of galaxy luminosity functions, the
star-formation main sequence and the bulk of the rest-frame
UV-optical-near infrared colour space. This suggests that the
aggregate star-formation history predicted by the model
successfully reproduces the trends in the observed galaxy
population.

{\color{black} We conclude this section by briefly noting how well
other hydrodynamical cosmological simulations reproduce the
observables that we have studied in this section. While rest-frame
luminosities and UV-optical-NIR colours have not been tested in
other similar simulations over a large redshift range as in this
study, the reproduction of rest-frame \emph{optical} colours and
$u$ to $K$-band luminosities in the \emph{local} Universe is found
to be generally good in such models \cite[e.g.][]{Trayford2015}.
In a similar vein, other simulations that are comparable to
Horizon-AGN generally show consistency with the observed star
formation main sequence \citep[e.g.][]{Furlong2015,Sparre2015},
although in some cases offsets of $\sim0.2-0.4$ dex are seen in
the normalization of the predicted sequences compared to the
observational ones \citep{Furlong2015}.}


\section{Stellar mass functions and the cosmic star formation
history} We proceed by comparing the predicted stellar mass
functions in Horizon-AGN to an array of observational data at
$0<z<6$ (Figure \ref{fig:mf}). The grey shaded region in this plot
indicates the Horizon-AGN predictions (with the width of the
region indicating Poisson uncertainties). The pink dashed line
indicates predictions from Horizon-noAGN, a twin simulation
without BH feedback. Not unexpectedly, low-mass galaxies are
largely unaffected by BH feedback. {\color{black}However,
agreement between theory and observation at the high-mass end of
the mass function depends strongly on implementing feedback from
BHs, confirming the results found in the literature
\citep[e.g.][]{Oppenheimer2008,Sales2010,McCarthy2010,Vogelsberger2014,Khandai2015,Sijacki2015,Crain2015}}.
The role of BHs is most important after the epoch of peak cosmic
star formation ($z \sim2$), as the BHs help keep star formation
rates low in massive galaxies (e.g. by maintaining the temperature
of their hot gas reservoirs) and prevent them from becoming too
massive. A forthcoming paper in this series (Beckmann et al. in
prep) will perform a detailed study of the role of AGN in
regulating the inflow of gas into galaxies and shaping the
evolving stellar mass function across cosmic time.

{\color{black}In a similar vein to the analyses presented above,
the predictions from Horizon-AGN show good consistency with the
slope and normalisation of observed mass functions within the
observational uncertainties.} The agreement is particularly good
for galaxies that are more massive than the knee of the mass
function at all epochs. However, two points of tension between
theory and observation are worth noting here: the overproduction
of galaxies at the low-mass end and the general underproduction of
galaxies regardless of galaxy stellar mass at $z>5$. Similarly to
what is seen in the luminosity-function analysis above,
Horizon-AGN tends to overproduce galaxies less massive than the
knee of the stellar mass function at low and intermediate redshift
{\color{black}($z<2$)}. While the predictions are consistent with
observations if uncertainties due to cosmic variance are also
taken into account (see the vertical error bars in Figure
\ref{fig:mf}), the systematic nature of the overproduction at all
epochs indicates that this disagreement may be due to missing
physics and that the sub-grid SN feedback prescription employed by
Horizon-AGN is not strong enough to regulate star formation in
smaller haloes, making the galaxies embedded in them too massive.

Indeed, recent work \citep{Kimm2015} has shown that a more
realistic treatment of the (inherently clumpy) interstellar medium
and the momentum injection from SN populations in the snowplough
phase \citep[see
also][]{Thornton1998,Hopkins2014,Geen2015,Kim2015,Martizzi2015}
can produce strong SN-driven outflows that regulate star formation
more efficiently than classical treatments. Alternatively, the
simulated galaxy stellar masses can be reduced if the star
formation efficiency is locally enhanced to the level that we
observe in star clusters ($\sim$ 10 \% per free fall time), as
clustered star formation drives stronger winds \citep{Agertz2015}
- enhanced efficiencies are indeed plausible, given the large
fraction of field stars that are thought to come from the
dissolution of star clusters \citep[see e.g.][]{Whitmore2007}.
Such improved prescriptions for star formation are able to bring
simulated galaxies in line with key observables, such as the
stellar-to-halo mass relation and the mass-metallicity relation.
While their implementation in Horizon-AGN is beyond the scope of
this particular paper, such revised recipes are likely to be a
solution to the disagreement observed at the low-mass end, and
will be explored in forthcoming papers.


\begin{figure}
$\begin{array}{c}
\includegraphics[width=3.5in]{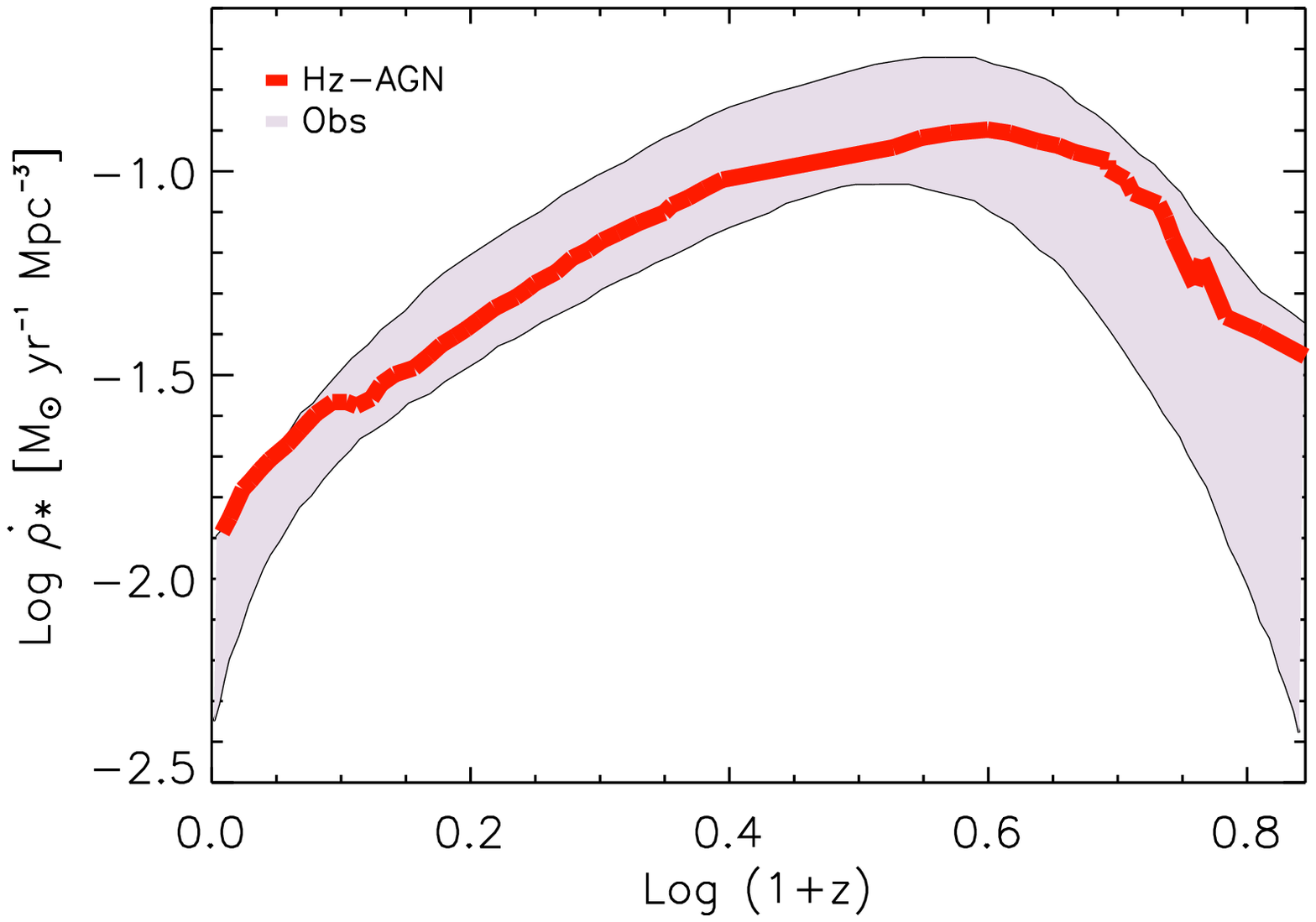}\\
\end{array}$
\caption{Comparison of the predicted cosmic star formation history
in Horizon-AGN to observational data \citep{Hopkins2006} in the
redshift range $0<z<6$. The red curve shows predictions from
Horizon-AGN while the grey shaded region indicates the parameter
space covered by the observational data.} \label{fig:cosmicsfh}
\end{figure}


We now turn to the second point of tension between theory and
observation. While the reproduction of the stellar mass function
is good across cosmic time, it appears to break down at $z\sim5$.
At this epoch, model galaxies appear to be less massive than their
observed counterparts across the entire mass range probed by our
study. We note here that the mass and spatial resolution of a
simulation can play an important role in influencing the predicted
stellar mass growth of galaxies \citep{Kimm2012}. Adopting a
higher resolution enables us to resolve smaller haloes in the
early universe ($z>5$), leading to earlier star formation. This
can result in an order of magnitude enhancement in the star
formation activity in typical galaxies at this epoch
\citep[e.g.][]{Rasera2006}. It is, therefore, likely that a
higher-resolution simulation can reduce the disagreement between
the theoretical and observed mass functions at this epoch, without
changing the baryonic physics currently implemented in
Horizon-AGN.

We proceed by comparing the cosmic star formation history (SFH)
predicted by Horizon-AGN to observations (Figure
\ref{fig:cosmicsfh}). The good reproduction of the luminosity and
stellar-mass functions, the star formation main sequence and
rest-frame colours already suggests that the simulation captures
the general trends in the evolution of galaxies over cosmic time.
This agreement is reflected and summarized in the comparison of
the cosmic star formation history. Not unexpectedly, Horizon-AGN
shows agreement with the observations, tracking both the shape and
normalization of the observational data \citep{Hopkins2006},
{\color{black}although we note that the predictions are close to
the upper bound of the parameter space defined by the
observational uncertainties at very low redshift ($z\lesssim
0.1$).}

{\color{black}We conclude this section by briefly discussing the
performance of similar models in reproducing the observed stellar
mass function and cosmic SFH. In the redshift range studied here,
most simulations that are similar to Horizon-AGN produce
reasonably good agreement with data
\citep[e.g.][]{Furlong2015,Genel2014,Khandai2015}, although a
detailed comparison between models is difficult, because all
models are typically not compared to the same observational
datasets at the same redshifts. However, similar patterns are seen
when most simulations are confronted with observational data.
While galaxies beyond the knee of the luminosity/stellar-mass
functions are well reproduced (due mainly to implementation of AGN
feedback, as noted above), models typically overshoot the data at
low stellar masses/luminosities at low and intermediate redshifts
($z<3$), with the level of the discrepancy varying with the model
in question. It is worth noting that the undershoot of the model
compared to data seen in Horizon-AGN at high redshift ($z>5$) is
also seen in other simulations
\citep[e.g.][]{Furlong2015,Genel2014}. Predictions made by these
models for the cosmic SFH are generally consistent with
observations - while all models reproduce the shape of the cosmic
SFH, small systematic offsets are seen in some comparisons
\citep[e.g.][]{Furlong2015}.}


\section{Summary}
We have compared the predictions of Horizon-AGN, a hydro-dynamical
cosmological simulation that uses an adaptive mesh refinement
code, to an extensive array of observational data across cosmic
time. Our study has focussed on confronting the predicted
evolution of luminosity functions, stellar-mass functions, the
star formation main sequence, rest-frame UV-optical-near infrared
colours and the cosmic star formation history to observational
data in the redshift range $0<z<6$ (around 95\% of the lifetime of
the Universe). These observables, which are functions of the
evolving stellar mass growth of the galaxy population, represent a
significant constraint on the methodologies employed by the
simulation. We note that, apart from choosing BH feedback
parameters that reproduce the local M$_{\rm BH}$ - $\sigma_*$
relations, the simulation is not otherwise calibrated to the local
Universe.

Since they are sensitive to the aggregate star formation history
of galaxies, comparing the simulation to these quantities
indicates how well Horizon-AGN captures the evolutionary trends of
observed galaxies over cosmic time, and thus its usefulness as a
tool for understanding galaxy evolution. Our analysis shows that
Horizon-AGN produces good agreement with the quantities mentioned
above across our redshift range of interest, from the present day
all the way to the epoch when the Universe was $\sim5$\% of its
present age. This indicates that model galaxies in the simulation
broadly reproduce the cosmic star formation history of their
counterparts in the real Universe.

Notwithstanding the reproduction of key observations by
Horizon-AGN, two main points of tension are worth noting. First,
the model tends to overproduce galaxies that are less massive than
the knee of the luminosity function at all epochs. While agreement
can be achieved by considering observational uncertainties due to
cosmic variance, it is possible that the SN feedback prescriptions
in the model do not sufficiently quench star formation in small
halos. More accurate modelling of the clumpy inter-stellar medium,
combined with higher star-formation efficiencies that correspond
to star-cluster formation may offer a solution to this
disagreement. Secondly, at very early epochs ($z \sim5$) the
predicted galaxy stellar masses are too low, across the mass range
of interest in our study. Higher-resolution simulations, that
resolve smaller haloes in the early Universe and their associated
star formation, are likely to reduce the disagreement between
theory and observation at these epochs.

Nevertheless, the good reproduction of the array of key
observations presented here indicates that the sub-grid recipes
that drive the baryonic evolution in the model are reasonably
accurate representations of the processes governing galaxy
evolution in the real Universe. Overall, we find that Horizon-AGN
offers an excellent tool, both for studying galaxy evolution over
cosmic time, and for making predictions for the next generation of
galaxy surveys.


\section*{Acknowledgements}
SK acknowledges a Senior Research Fellowship from Worcester
College Oxford. CL is supported by the ILP LABEX (under reference
ANR-10-LABX-63 and ANR-11-IDEX-0004-02). TK is supported by the
ERC Advanced Grant 320596 `The Emergence of Structure during the
Epoch of Reionization'. JD and AS acknowledge funding support from
Adrian Beecroft, the Oxford Martin School and the STFC. This
research has used the DiRAC facility, jointly funded by the STFC
and the Large Facilities Capital Fund of BIS, and has been
partially supported by grant Spin(e) ANR-13-BS05-0005 of the
French ANR. This work was granted access to the HPC resources of
CINES under the allocations 2013047012, 2014047012 and 2015047012
made by GENCI. We thank Stephane Rouberol for running and
maintaining the Horizon cluster hosted by the Institut
d'Astrophysique de Paris and the COSMOS2015 team for allowing us
to use their data ahead of publication. This work is part of the
Horizon-UK project.


\nocite{Kaviraj2007a,Kaviraj2007b,Ilbert2005,Fontana2004,Bielby2012,Tomczak2014,Borch2006,Moustakas2013,Pozzetti2007,Volonteri2016,Santini2012,Marchesini2009,Perezgonzalez2008,Caputi2015,Song2015,Lee2012,Gonzalez2011,Driver2012,Jones2006,Eke2005,Arnouts2007,Driver2012,Bell2003,Loveday2012,Ilbert2005,Dahlen2005,Welker2015a,Welker2015b,Dubois2014,Codis2015,Sutherland1993,Chisari2016,Dopita2005,Groves2008}


\bibliographystyle{mn2e}
\bibliography{references}


\section*{Appendix A}
\begin{figure}
$\begin{array}{c}
\includegraphics[width=3.5in]{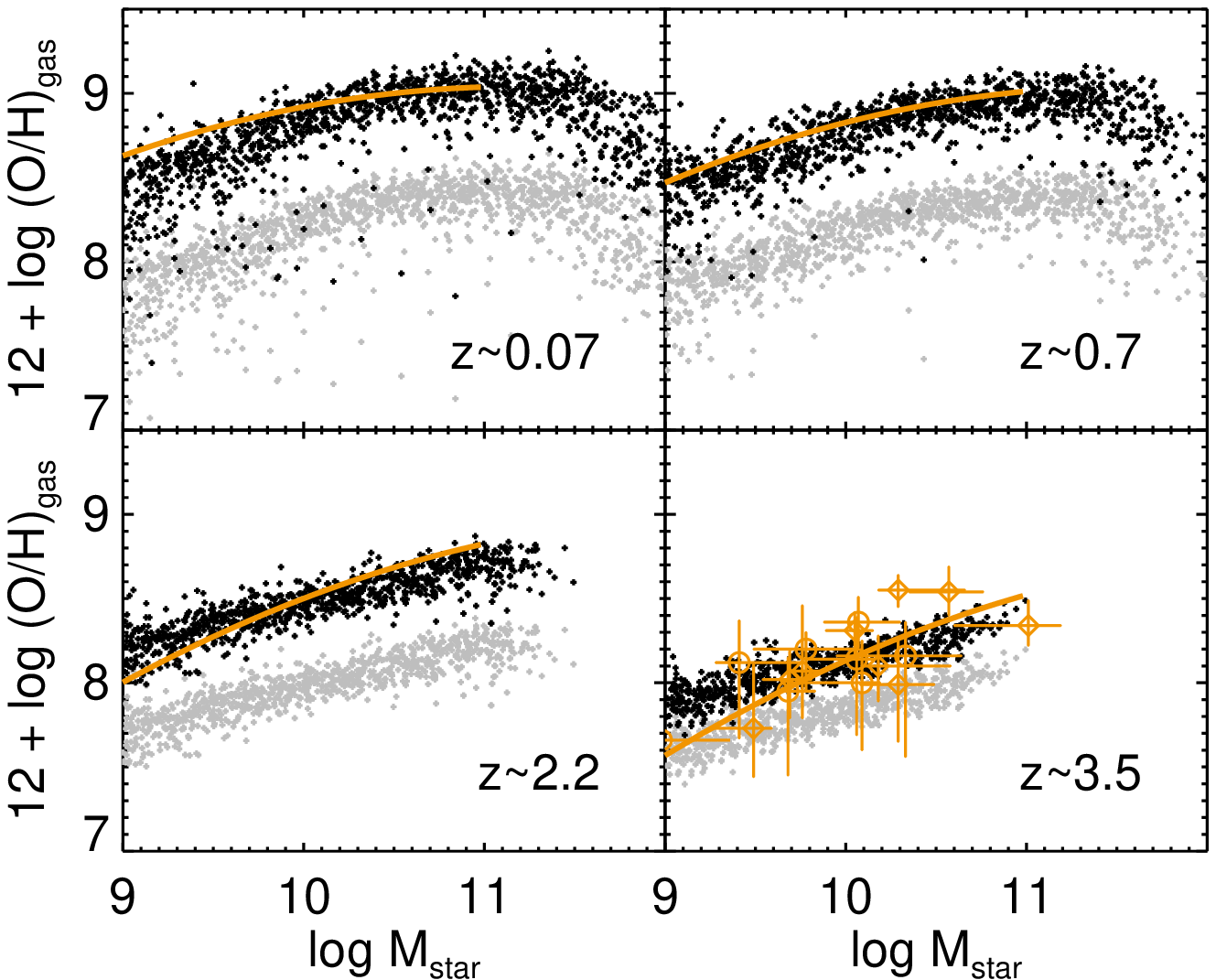}\\
\end{array}$
\caption{{\color{black}Calibration of the predicted mass-gas phase
metallicity relations to osbervational data. The mass-gas phase
metallicity relations predicted by Horizon-AGN are shown in grey.
The observational data \citep{Maiolino2008,Mannucci2009} to which
we perform the calibrations to correct the simulated metallicities
are shown in orange. Best fits to the observational datasets are
shown using the orange lines. For the $z\sim3.5$ dataset we also
show the individual points to give an indication of the typical
scatter around the best-fit lines (we only show one set of points
for clarity and because the scatter is similar at all epochs). The
corrected metallicities that are used for the analysis in this
study are shown using the black points.}}
\label{fig:metallicity_calibration}
\end{figure}

{\color{black}As described in Section 3, the simulated gas-phase
metallicities in Horizon-AGN are under-estimated compared to
observations due to a delayed enrichment of star-forming clouds,
which is essentially an artifact of the (relatively low)
resolution of the simulation. To correct this, the gas-phase
metallicities are multiplied by a redshift-dependent
renormalisation factor ($f_{\rm no}$) that brings the simulated
metallicities in agreement with the observed mass-gas phase
metallicity relations at $z=0$, 0.7, 2.5 and 3.5
\citep{Maiolino2008,Mannucci2009}, where $f_{\rm no}=4.08 - 0.21z
- 0.11z^2$. As noted before, the observed mass-metallicity
relations are calculated using strong line diagnostics (e.g.
[OIII]5007/H$\beta$, [OIII]5007/[OII]3727, [NeIII]3870/[OII]3727).

In Figure \ref{fig:metallicity_calibration} we show the mass-gas
phase metallicity relations predicted by the simulation using the
grey points. The observational data
\citep{Maiolino2008,Mannucci2009} to which we perform the
calibrations to correct the simulated metallicities are shown in
orange. Best fits to the observational datasets are shown using
the orange lines. For the $z\sim3.5$ dataset, we also show the
individual points to give an indication of the typical scatter
around the best-fit lines (we only show one set of points for
clarity, and because the scatter is similar at all epochs). The
corrected metallicities that are used in calculating the
properties of simulated galaxies in the model are shown using the
black points. Since the shape of predicted mass-gas phase
metallicity relation is reasonably consistent with the shape of
their observed counterparts at redshifts where data is available,
we do not attempt to make the calibration stellar mass-dependent -
the normalization is simply adjusted as a function of redshift,
following $f_{\rm no}$.}

\end{document}